\documentclass[sigconf, pbalance]{acmart}

\copyrightyear{2026}
\acmYear{2026}
\setcopyright{cc}
\setcctype{by}
\acmConference[HASP '26]{International Workshop on Hardware and Architectural Support for Security and Privacy 2026}{October 31-November 01, 2026}{Athens, Greece}
\acmBooktitle{International Workshop on Hardware and Architectural Support for Security and Privacy 2026 (HASP '26), October 31-November 01, 2026, Athens, Greece}
\acmDOI{10.1145/3857307.3857308}
\acmISBN{979-8-4007-2964-5/2026/10}

\usepackage{tikz}
\usetikzlibrary{matrix, positioning, arrows.meta, calc, fit, backgrounds, patterns, patterns.meta, shapes.geometric, shapes.multipart}

\usepackage{prettyref}
\newrefformat{sec}{Section~\ref{#1}}
\newrefformat{fig}{Figure~\ref{#1}}
\newrefformat{app}{Appendix~\ref{#1}}

\newcommand{\bld}{\mathrm{Bld}}
\newcommand{\aud}{\mathrm{Aud}}
\newcommand{\aux}{\mathit{Aux}}
\newcommand{\conf}{\mathit{Conf}}
\newcommand{\verd}{\mathit{Verdict}}

\newcommand{\meas}{\operatorname{meas}}
\newcommand{\AR}{\operatorname{AR}}

\newcommand{\MRENCLAVE}{\texttt{MRENCLAVE}}
\newcommand{\MEASUREMENT}{\texttt{MEASUREMENT}}
\newcommand{\MRTD}{\texttt{MRTD}}
\newcommand{\HOSTDATA}{\texttt{HOST\_DATA}}
\newcommand{\MRCONFIGID}{\texttt{MR\_CONFIG\_ID}}
\newcommand{\REPORTDATA}{\texttt{REPORT\_DATA}}

\tikzset{
  rbox/.style    = {draw, rounded corners=2pt, align=center, inner sep=3pt, font=\footnotesize, minimum height=7mm},
  arr/.style     = {-Stealth, semithick},
  dasharr/.style = {-Stealth, semithick, dashed},
}

\definecolor{cudYellow}{HTML}{F0E442}
\definecolor{cudSky}{HTML}{56B4E9}
\definecolor{cudVermillion}{HTML}{D55E00}
\definecolor{cudBlue}{HTML}{0072B2}

\newcommand{\IconScope}[2]{\begin{scope}[shift={(path picture bounding box.center)},x=#2,y=#2,#1,line width=0.45pt,line join=round,line cap=round]}
\newcommand{\EndIcon}{\end{scope}}
\newcommand{\PicCode}[2]{\IconScope{#1}{#2}%
  \draw (-3.6,-4.4)--(-3.6,4.4)--(1.4,4.4)--(3.6,2.2)--(3.6,-4.4)--cycle;
  \draw (1.4,4.4)--(1.4,2.2)--(3.6,2.2);
  \draw[line width=0.65pt] (-0.95,1.35)--(-2.25,0.0)--(-0.95,-1.35);
  \draw[line width=0.65pt] ( 0.95,1.35)--( 2.25,0.0)--( 0.95,-1.35);
  \draw[line width=0.65pt] (0.65,-1.9)--(-0.65,1.9);
\EndIcon}
\newcommand{\PicConf}[2]{\IconScope{#1}{#2}%
  \foreach \y/\x in {2.0/1.0, 0/-0.9, -2.0/0.6}{%
    \draw[line width=0.5pt] (-2.8,\y)--(2.8,\y);
    \fill (\x,\y) circle (0.6);
    \draw[line width=0.5pt] (\x,\y) circle (0.6);}
\EndIcon}
\newcommand{\PicDB}[2]{\IconScope{#1}{#2}%
  \draw (-2.4,2.6) -- (-2.4,-2.6);
  \draw (2.4,2.6) -- (2.4,-2.6);
  \draw (-2.4,-2.6) arc (180:360:2.4 and 0.9);
  \draw (0,2.6) ellipse (2.4 and 0.9);
  \draw (0,0.2) ellipse (2.4 and 0.9);
\EndIcon}
\newcommand{\PicCube}[2]{\IconScope{#1}{#2}%
  \draw (-2.7,-1.7)--(0,-3.3)--(2.7,-1.7)--(2.7,1.7)--(0,3.3)--(-2.7,1.7)--cycle;
  \draw (-2.7,-1.7)--(0,-0.1)--(2.7,-1.7);
  \draw (0,-0.1)--(0,3.3);
\EndIcon}
\newcommand{\PicReport}[2]{\IconScope{#1}{#2}%
  \draw (-3.4,-4.4)--(-3.4,4.6)--(1.4,4.6)--(3.4,3.0)--(3.4,-1.0);
  \draw (-3.4,-4.4)--(0.2,-4.4);
  \draw (1.4,4.6)--(1.4,3.0)--(3.4,3.0);
  \foreach \y in {2.4,1.2,0.0}{\draw[line width=0.32pt] (-2.4,\y)--(2.2,\y);}
  \foreach \a in {0,30,...,330}{\draw[line width=0.3pt] (1.7,-2.7)+(\a:1.25)--+(\a:1.85);}
  \draw[line width=0.35pt] (1.7,-2.7) circle (1.25);
  \draw[line width=0.45pt] (1.05,-2.8)--(1.55,-3.4)--(2.4,-2.0);
  \draw[line width=0.32pt] (1.15,-3.8)--(0.75,-5.2)--(1.55,-4.55);
  \draw[line width=0.32pt] (2.25,-3.8)--(2.65,-5.2)--(1.85,-4.55);
\EndIcon}
\newcommand{\PicChip}[2]{\IconScope{#1}{#2}%
  \draw (-2.7,-2.7) rectangle (2.7,2.7);
  \foreach \x in {-1.3,0,1.3}{%
    \draw (\x,2.7)--(\x,3.7); \draw (\x,-2.7)--(\x,-3.7);
    \draw (2.7,\x)--(3.7,\x);  \draw (-2.7,\x)--(-3.7,\x);}
  \draw (-1.2,-1.4) rectangle (1.2,0.6);
  \draw (-0.7,0.6) arc (180:0:0.7);
  \draw[line width=0.5pt] (0,-0.1)--(0,-0.7);
\EndIcon}
\newcommand{\PicServer}[2]{\IconScope{#1}{#2}%
  \draw (-3,-3.8) rectangle (3,-1.2);
  \draw (-3,-0.4) rectangle (3,2.2);
  \fill (-2.1,-2.5) circle (0.28);
  \fill (-2.1,0.9) circle (0.28);
  \draw[line width=0.32pt] (-1.2,-2.5)--(2.2,-2.5);
  \draw[line width=0.32pt] (-1.2,0.9)--(2.2,0.9);
\EndIcon}
\newcommand{\PicVerify}[2]{\IconScope{#1}{#2}%
  \draw[line width=0.5pt] (-0.7,0.9) circle (2.1);
  \draw[line width=0.7pt] (0.75,-0.55)--(3.0,-2.8);
  \draw[line width=0.5pt] (-1.8,1.0)--(-1.05,0.1)--(0.4,1.7);
\EndIcon}
\newcommand{\PicDev}[2]{\IconScope{#1}{#2}%
  \draw (0,1.8) circle (1.5);
  \draw (-2.9,-3.4) arc (180:0:2.9 and 3.4) -- cycle;
\EndIcon}
\newcommand{\PicContainer}[2]{\IconScope{#1}{#2}%
  \draw (-3.4,-2.6) rectangle (3.4,2.6);
  \draw[line width=0.35pt] (-3.4,1.6)--(3.4,1.6);
  \foreach \x in {-2.2,-0.8,0.6,2.0}{\draw[line width=0.35pt] (\x,-2.6)--(\x,1.6);}
\EndIcon}
\newcommand{\PicVerdict}[2]{\IconScope{#1}{#2}%
  \draw (-3.4,-4.4)--(-3.4,4.6)--(1.4,4.6)--(3.4,3.0)--(3.4,-4.4)--cycle;
  \draw (1.4,4.6)--(1.4,3.0)--(3.4,3.0);
  \foreach \y in {2.4,1.2,0.0,-1.2}{\draw[line width=0.32pt] (-2.4,\y)--(2.2,\y);}
  \draw[line width=0.32pt] (-2.4,-2.4)--(0.4,-2.4);
\EndIcon}
\newcommand{\PicHash}[2]{\IconScope{#1}{#2}%
  \draw[line width=0.6pt] (-2.6,1.1)--(2.6,1.1);
  \draw[line width=0.6pt] (-2.6,-1.1)--(2.6,-1.1);
  \draw[line width=0.6pt] (-0.9,2.6)--(-1.7,-2.6);
  \draw[line width=0.6pt] (1.7,2.6)--(0.9,-2.6);
\EndIcon}
\newcommand{\PicSeal}[2]{\IconScope{#1}{#2}%
  \foreach \a in {0,30,...,330}{\draw[line width=0.32pt] (0,0.9)+(\a:2.0)--+(\a:2.7);}
  \draw[line width=0.4pt] (0,0.9) circle (2.0);
  \draw[line width=0.4pt] (0,0.9) circle (1.15);
  \draw[line width=0.55pt] (-0.7,0.9)--(-0.15,0.25)--(0.85,1.5);
  \draw[line width=0.4pt] (-0.9,-0.8)--(-1.35,-3.0)--(-0.2,-2.1);
  \draw[line width=0.4pt] (0.9,-0.8)--(1.35,-3.0)--(0.2,-2.1);
\EndIcon}

\begin{document}

\title{Attestable Audit: Property-Based Attestation for Proprietary Workloads without Artifact Disclosure}

\author{Takuma Imamura}
\correspondingauthor
\affiliation{
  \institution{Confidential Computing Lab, Acompany Co., Ltd.}
  \city{Nagoya}
  \country{Japan}}
\orcid{0000-0001-6491-9137}
\email{takuma.imamura@acompany-ac.com}

\begin{abstract}

Trusted Execution Environments (TEEs) prove the integrity of code and data to a remote peer through remote attestation: the TEE issues an attestation report by signing, with a hardware-bound key, its configuration information, including measurements (cryptographic digests) of the code loaded inside it. A verifier appraises the report by matching these measurements against reference values. When the measured artifact is proprietary, however, the verifier cannot derive the reference values itself: it can only confirm that \emph{something} with a given measurement is running inside the TEE---not that the program behaves as expected---forcing full trust in the reference value provider. We propose an architecture that closes this gap without disclosing the artifact, composing two TEE workloads: \emph{attestable build} (Hugenroth \textit{et al.}, 2025), which binds a source code digest to a build artifact measurement, and \emph{attestable audit} (proposed in this paper), which binds the same source code digest to the verdicts of automated audits---fuzzing, static analysis, AI code auditing, formal methods---executed inside a TEE. Both are instances of Confidential Computing Proofs: TEE attestation viewed as a hardware-backed zero-knowledge proof. Chaining the two certificates lets the verifier conclude that the running artifact was built from source code satisfying the audited properties, \emph{without disclosing either the source code or the build artifact}. A proof-of-concept implementation on Intel SGX with the Gramine Library OS, running the Bandit security analyzer on Python code, demonstrates the practicality of the approach.

\end{abstract}




\maketitle

% ---------------------------------------------------------------------
%  Introduction
% ---------------------------------------------------------------------
\section{Introduction}
\label{sec:intro}

\subsection{Background}
Remote attestation (RA) for Trusted Execution Environments (TEEs) is a protocol by which a TEE proves the integrity of its code and data to a remote peer. Throughout this paper we follow the terminology of the RATS architecture \citep{rfc9334}, which defines the following roles. The \emph{attester} is the platform under attestation; it issues \emph{evidence} containing claims about its own state. The \emph{verifier} appraises the evidence and sends the \emph{attestation results} to the \emph{relying party}, the party that decides, based on those results, whether to interact with the attester. In this appraisal, the verifier validates the authenticity of the evidence using \emph{endorsements} (such as certificate chains and certificate revocation lists) issued by an \emph{endorser} (typically the hardware vendor), and matches the measurements contained in the evidence against \emph{reference values} supplied by a \emph{reference value provider} (typically the software vendor).

The basic flow of RA is as follows. The attester measures cryptographic hash values of the code loaded into its isolated execution context, and issues as evidence an \emph{attestation report}: the system's configuration information, including these \emph{measurements}, signed with an \emph{attestation key} bound to the TEE platform. The verifier validates the authenticity of the attestation report using the endorsements, and then matches the measurements in the report against the reference values, thereby confirming that the expected code is running inside the TEE.

This `measurement-and-comparison' pattern of RA appears consistently throughout the TEE ecosystem in several specific forms.

The most basic form of this pattern is the RA for the initial TEE state. In Intel SGX, the digest of the loaded enclave image {\MRENCLAVE} is measured at load time and bound into the attestation report (SGX quote) \citep{Cos16}. In Confidential VMs (CVMs) such as AMD SEV-SNP and Intel TDX, the digest of the VM's initial memory state ({\MEASUREMENT} and {\MRTD}, respectively) is measured at VM launch and bound into the attestation report (SNP report and TD quote, respectively). The verifier checks these load-time measurements against reference values to confirm that the expected enclave or VM image is running.

Another, more complex pattern can be observed in CVM's runtime attestation. While the above measurements cover only the initial state of the TEE, runtime integrity in CVMs is typically handled by Kernel's runtime integrity protection system, such as the Linux Integrity Measurement Architecture (IMA) \citep{SZJD04}. IMA is a Linux kernel subsystem that measures and logs hash digests of files accessed inside the VM according to a loaded IMA policy. To make the measurement log itself attestable, IMA extends each entry into an append-only register that is isolated from both the guest OS and the hypervisor: a Platform Configuration Register (PCR) of a vTPM residing inside the guest VM (e.g. VMPL 0 for SEV-SNP \citep{NarEtAl23}), or a Runtime Measurement Register (RTMR) provided by Intel TDX \citep{tdvf23}. After verifying the authenticity of the IMA log against this register, the verifier appraises each entry---consisting of a file path and a file hash---according to some policy. For example, the policy may allow or deny specific combinations of file paths (or patterns) and hashes (or hash lists, such as the set of hashes corresponding to an acceptable version range of a binary). Denial policies are particularly useful for excluding library versions known to contain severe vulnerabilities or for blocking known malware.

Yet another form of the pattern appears in the RA for Confidential Containers (CoCo). CoCo \citep{coco} is a container running inside a CVM. The host configures the hash digest of a container policy---which itself contains the digest of the permitted container images---in a configurable register (e.g. {\HOSTDATA} for SEV-SNP or {\MRCONFIGID} for TDX) at guest VM launch, and passes the corresponding policy to a policy engine inside the guest VM. The policy engine verifies the authenticity of the received policy by comparing its hash against the policy measurement field (e.g. {\HOSTDATA} or {\MRCONFIGID}) in the attestation report, and only then launches the container according to the policy. The verifier validates the entire chain by

\begin{enumerate}
\item comparing the launch measurement (e.g. {\MEASUREMENT} or {\MRTD}) in the attestation report against its reference value to confirm that the expected container runtime and policy engine are running,
\item comparing the policy measurement field (e.g. {\HOSTDATA} or {\MRCONFIGID}) in the attestation report against the hash of the expected container policy to confirm that the expected policy is loaded inside the CVM, and
\item comparing the container image digest embedded in the policy against the reference value to confirm that the expected container is running.
\end{enumerate}

In each of these cases, the attester provides the verifier with evidence that includes digests of its own program artifacts---VM initial images; kernel modules, libraries, and executables inside the guest OS; container images; among others---and the verifier validates this evidence by comparing the digests against reference values.

When the program in question is open source (and, strictly speaking, its build is reproducible \citep{LZ22}), the verifier can prepare these reference values itself by building from source and computing the resulting digests. Even when the source is not open, the verifier can still derive reference values independently as long as the executable or container image itself is publicly available.

\begin{figure*}[tb]
\centering

\begin{tikzpicture}[font=\footnotesize,
  tile/.style    = {draw, rounded corners=2pt, align=center, inner sep=2pt, thick},
  trustT/.style  = {tile, fill=green!12, draw=green!65!black, minimum width=14mm, minimum height=12mm},
  untrustT/.style= {tile, fill=red!12,  draw=red!70!black,   minimum width=14mm, minimum height=12mm},
  alab/.style    = {font=\scriptsize, align=center, inner sep=2.5pt},
  rlab/.style    = {font=\scriptsize, align=left, inner sep=2.5pt},
  tlab/.style    = {font=\scriptsize\itshape},
  arl/.style     = {font=\scriptsize\itshape},
  r@tee/.style = {trustT,  path picture={\PicChip{green!35!black}{0.72mm}}},
  r@cc/.style  = {trustT,  path picture={\PicContainer{green!35!black}{0.72mm}}},
  r@ima/.style = {trustT,  path picture={\PicServer{green!35!black}{0.72mm}}},
  r@ver/.style = {trustT,  path picture={\PicVerify{green!35!black}{0.8mm}}},
  r@end/.style = {trustT,  path picture={\PicSeal{green!35!black}{0.8mm}}},
  r@rvp/.style = {untrustT,path picture={\PicDB{red!65!black}{0.8mm}}},
  r@rp/.style  = {trustT,  path picture={\PicDev{green!35!black}{0.8mm}}},
  arr/.style   = {-{Stealth[length=2.2mm]}, line width=0.6pt, gray!55},
]

\node[r@ver] (v)   at (0,0)     {};
\node[r@rvp] (rvp) at (2.5,1.8)  {};
\node[r@end] (end) at (2.5,-1.8) {};
\node[r@rp]  (rp)  at (-4.4,0)  {};

\node[r@tee] (tee) at (5.4,1.8)  {};
\node[r@cc]  (cc)  at (5.4,0)    {};
\node[r@ima] (ima) at (5.4,-1.8) {};
\node[rlab, right=1.5mm of tee, name=teel] {TEE app/VM\\{\scriptsize(SGX, SEV-SNP, TDX)}};
\node[rlab, right=1.5mm of cc,  name=ccl]  {Confidential Container};
\node[rlab, right=1.5mm of ima, name=imal] {Linux OS with\\TPM-backed IMA};
\node[fit=(tee)(ima)(teel)(imal), draw, dashed, inner sep=4pt, rounded corners,
      label={[font=\footnotesize]below:Attesters (hash-emitting platforms)}] (attbox) {};

\draw[arr] (rvp.west) -| (v.north);
\draw[arr] (end.west) -| (v.south);
\draw[arr] (attbox.west) -- (v.east);
\draw[arr] (v.west) -- (rp.east);
\node[arl, anchor=center] at (2.5, 0.3) {evidence};
\node[arl, anchor=center] at (-2, 0.3) {attestation results};
\node[arl, anchor=east] at (-0.10,1.5) {reference values};
\node[arl, anchor=east] at (-0.10,-1.5) {endorsements};

\node[alab, below=0.5mm of rvp] {Reference Value\\Provider};
\node[tlab, red!70!black, above=0.5mm of rvp] {trusted / untrusted};
\node[alab, below=0.5mm of end] {Endorser};
\node[tlab, green!50!black, above=0.5mm of end] {trusted};
\node[alab, below=0.5mm of rp] {Relying Party};
\node[alab, below=0.5mm of v] {Verifier};
\node[tlab, green!50!black, above=0.5mm of v] {trusted};
\node[tlab, green!50!black, above=0.5mm of attbox] {attestable};
\end{tikzpicture}

\caption{RATS role model \citep{rfc9334}. The relying party trusts the verifier and the endorser. In a typical proprietary scenario, the relying party also trusts the reference value provider. Our goal is to remove that trust.}
\Description{RATS role model diagram. Attesters (TEE apps or VMs, confidential containers, and Linux hosts with TPM-backed IMA) send evidence to a trusted verifier. The reference value provider, marked as trusted or untrusted, supplies reference values to the verifier; a trusted endorser supplies endorsements. The verifier sends attestation results to the relying party.}
\label{fig:rats}

\end{figure*}
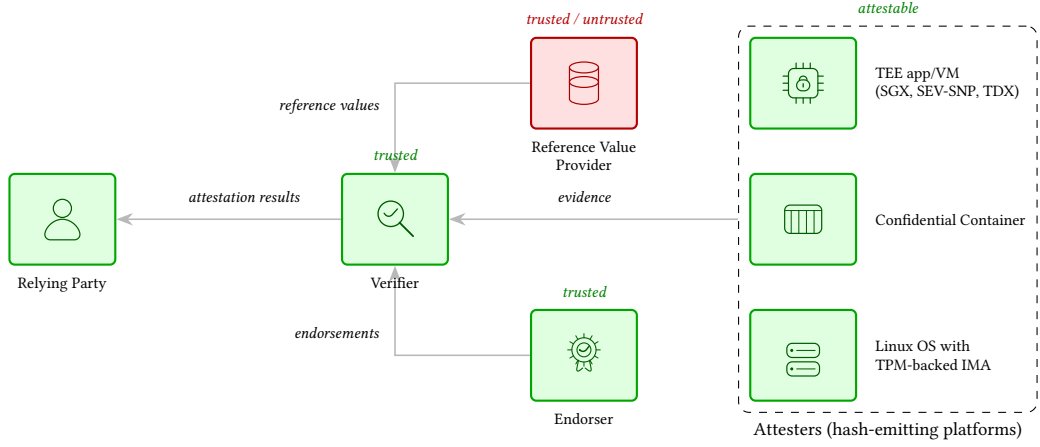

The situation differs fundamentally, however, when the program is closed and proprietary. In this case, the verifier cannot derive reference values on its own. Instead, a reference value provider distributes only the reference values to the verifier. The verifier can then confirm that ``something with this digest is running inside the TEE,'' but nothing more: it cannot determine what program is actually running, nor whether the program has any particular property of interest (e.g. the absence of a backdoor). This is essentially the same problem pointed out by Sadeghi and St{\"u}ble in 2004 \citep{SS04}: a binary's digest is just an identifier, not a semantic guarantee. In this case, the relying party is forced to place \emph{full trust in the reference value provider} in everything beyond the bare hash digest (see \prettyref{fig:rats}).

\subsection{Our contribution}

In this paper, we propose an architecture that closes this gap in remote attestation without disclosing the pre-hashed data. Our approach builds on two mechanisms.

\emph{Attestable build}, introduced by Hugenroth \emph{et al.} \citep{HLMB25}, builds source code inside an attestable isolated environment, such as a TEE\footnote{The paper \citep{HLMB25} primarily uses AWS Nitro Enclaves. Although Nitro Enclaves are attestable in the same way as TEEs, they differ from TEEs in that they rely on the hypervisor.}, to cryptographically bind the source code digest to the build artifact digest.

\emph{Attestable audit}, proposed in this paper, performs automated audits on source code inside a TEE to cryptographically bind the source code digest to properties of the code. Here, automated audit covers fuzzing, Valgrind \citep{Valgrind}, taint analysis, AI code auditing, and formal verification, among others.

These two mechanisms yield a chain from the build artifact digest---which is what a remote verifier observes in the attestation evidence of a running TEE---to verifiable properties of the source code, without disclosing either the source code or the build artifacts. Conceptually, our approach is grounded in the observation that remote attestation can be considered as a hardware-backed zero-knowledge proof (ZKP); see \citep{RusEtAl24} for further discussion.

To demonstrate the feasibility of attestable audit, we provide a proof-of-concept implementation on Intel SGX using the Gramine LibOS \citep{TPV17,gramine}. The PoC takes a Python script as input and runs Bandit \citep{bandit}, a static security analyzer, inside the SGX enclave; the resulting audit report is bound to the source code digest via the enclave's attestation report.

% ---------------------------------------------------------------------
%  System Model and Trust Assumptions
% ---------------------------------------------------------------------
\section{System model and trust assumptions}
\label{sec:model}

\emph{Parties.} We follow the standard RATS role model \citep{rfc9334} (see \prettyref{fig:rats}): the relying party trusts the verifier and the endorser. The attester may be any RATS-compliant platform---SGX enclaves, CVMs, TPM-backed systems, AWS Nitro Enclaves, etc.---and its trust assumptions are inherited from the underlying technology. The \emph{developer} of the attester's proprietary code acts as the reference value provider, and we use the two terms interchangeably.

\emph{Adversary model.} Our key departure from the standard model is that the relying party does \emph{not} trust the reference value provider: the adversary is a malicious developer who may present arbitrary---forged, stale, or mix-and-matched---reference values, certificates, and code, in order to make the relying party accept an attester that lacks the claimed properties. Conversely, the developer regards the verifier as honest-but-curious: it follows the protocol but attempts to learn the proprietary artifact. As usual in the TEE setting, the hardware root of trust and the chip vendor's PKI are trusted, whereas the host OS, the hypervisor, and all software outside a TEE are untrusted.

\emph{Security goals.} The architecture must provide (P1)~\emph{binding}: if the verifier accepts, the attester runs a build artifact whose source code actually produced the claimed audit verdicts under the disclosed audit configuration; and (P2)~\emph{non-disclosure}: the verifier learns nothing about the proprietary artifact beyond its digest and the disclosed audit configuration and verdicts.

To remove the trust in the reference value provider, we delegate the production of reference values and audit results to a pair of dedicated TEE applications---an \emph{attestable builder} performing attestable build and an \emph{attestable auditor} performing attestable audit---under the threat model above. For this delegation to fully remove the trust, the behaviour of the attestable builder and auditor must themselves be independently verifiable: \emph{their measurements must correspond to publicly inspectable artifacts,} i.e. publicly available build artifacts and/or reproducibly buildable source code. Otherwise, trust would simply shift to whoever supplies the reference measurements of the attestable builder and auditor. It is important to note that this requirement applies only to the attestable builder and auditor; the \emph{audited artifact itself remains undisclosed}, which is precisely the property our architecture is designed to preserve.

\emph{Out of scope.} Side-channel and physical attacks against the TEE platforms are outside the protocol analysis; we discuss the practical impact of low-cost physical attacks, and the hosting considerations they motivate, in \prettyref{sec:discussion}. We also take the semantic quality of the audit tools as given: what an audit verdict is worth is decided by the relying party's policy, not by our construction.

% ---------------------------------------------------------------------
%  Architecture
% ---------------------------------------------------------------------
\section{Architecture}
\label{sec:arch}

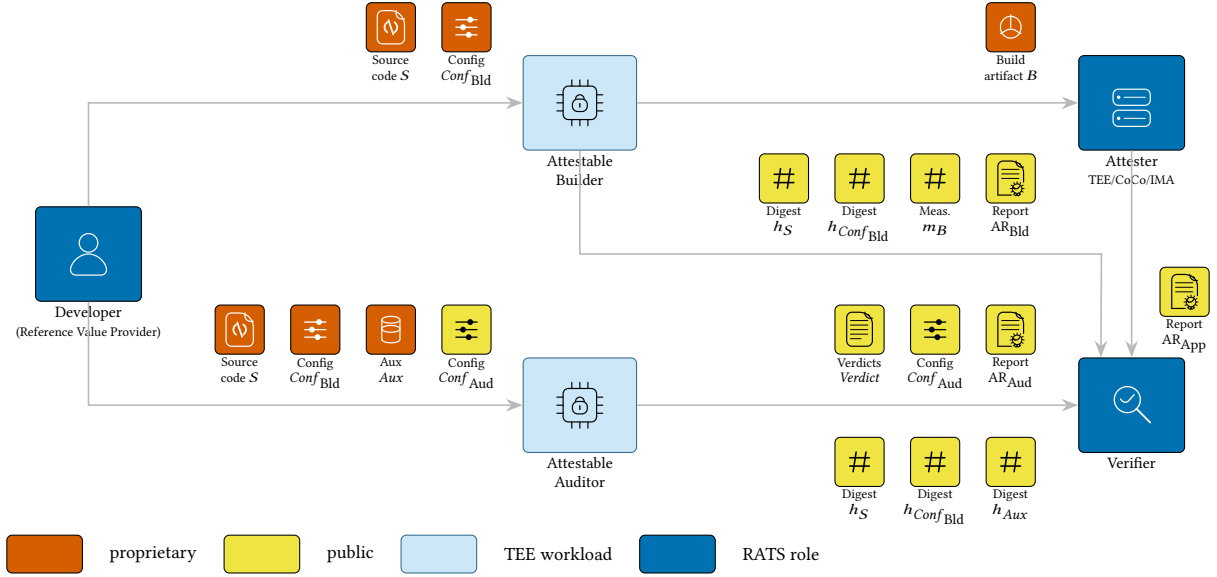
\begin{figure*}[tb]
\centering

\begin{tikzpicture}[
    font=\footnotesize,
    >={Stealth[length=2.2mm,width=1.8mm]},
    tile/.style  = {draw, rounded corners=2pt, align=center, inner sep=2pt},
    actor/.style = {tile, minimum width=14mm, minimum height=12.5mm},
    dev/.style   = {actor, fill=cudBlue, path picture={\PicDev{white}{0.82mm}}},
    encl/.style  = {actor, fill=cudSky!30, draw=cudSky!60!black, minimum width=15mm, path picture={\PicChip{black}{0.82mm}}},
    att/.style   = {actor, fill=cudBlue, path picture={\PicServer{white}{0.82mm}}},
    ver/.style   = {actor, fill=cudBlue, path picture={\PicVerify{white}{0.82mm}}},
    data/.style  = {tile, minimum width=6.5mm, minimum height=6.5mm, inner sep=1pt},
    d@code/.style = {data, fill=cudVermillion, path picture={\PicCode{white}{0.5mm}}},
    d@conf/.style = {data, fill=cudVermillion, path picture={\PicConf{white}{0.5mm}}},
    d@aux/.style  = {data, fill=cudVermillion, path picture={\PicDB{white}{0.5mm}}},
    d@caud/.style = {data, fill=cudYellow,     path picture={\PicConf{black}{0.5mm}}},
    d@bin/.style  = {data, fill=cudVermillion, path picture={\PicCube{white}{0.5mm}}},
    d@rep/.style  = {data, fill=cudYellow,     path picture={\PicReport{black}{0.5mm}}},
    d@hash/.style = {data, fill=cudYellow,     path picture={\PicHash{black}{0.5mm}}},
    d@verd/.style = {data, fill=cudYellow,     path picture={\PicVerdict{black}{0.5mm}}},
    albl/.style  = {font=\scriptsize, align=center, inner sep=1.5pt},
    dlbl/.style  = {font=\tiny, align=center, inner sep=0.8pt},
    arr/.style   = {-{Stealth[length=2.2mm]}, line width=0.6pt, gray!55},
]

\node[dev,  label={[albl]below:Developer\\{\tiny (Reference Value Provider)}}] (dev) at (0,0) {};
\node[encl, label={[albl]below:Attestable\\Builder}] (be) at (6.5,2) {};
\node[encl, label={[albl]below:Attestable\\Auditor}] (ae) at (6.5,-2) {};
\node[att,  label={[albl]below:Attester\\\tiny TEE/CoCo/IMA}] (dep) at (13.8,2) {};
\node[ver,  label={[albl]below:Verifier}] (v) at (13.8,-2){};

\draw[arr] (dev.north) |- (be.west);
\draw[arr] (dev.south) |- (ae.west);
\draw[arr] (be.east) -- (dep.west);
\draw[arr] (be.south) -- (6.5, 0) -- (13.4, 0) -- ([xshift=0.3cm]v.north west);
\draw[arr] (ae.east) -- (v.west);
\draw[arr] (dep.south) -- (v.north);

\node[d@code, label={[dlbl]below:Source\\code $S$}] at (4.0,3) {};
\node[d@conf, label={[dlbl]below:Config\\$\conf_{\bld}$}] at (5,3) {};
\node[d@code, label={[dlbl]below:Source\\code $S$}] at (2,-1) {};
\node[d@conf, label={[dlbl]below:Config\\$\conf_{\bld}$}] at (3,-1) {};
\node[d@aux,  label={[dlbl]below:Aux\\$\aux$}] at (4,-1) {};
\node[d@caud, label={[dlbl]below:Config\\$\conf_{\aud}$}] at (5,-1) {};

\node[d@bin, label={[dlbl]below:Build\\artifact $B$}] at (12.2,3) {};
\node[d@hash, label={[dlbl]below:Digest\\$h_{S}$}] at (9.2,1) {};
\node[d@hash, label={[dlbl]below:Digest\\$h_{\conf_{\bld}}$}] at (10.2,1) {};
\node[d@hash, label={[dlbl]below:Meas.\\$m_{B}$}] at (11.2,1) {};
\node[d@rep, label={[dlbl]below:Report\\$\AR_{\bld}$}] at (12.2, 1) {};
\node[d@verd, label={[dlbl]below:Verdicts\\$\verd$}] at (10.2,-1) {};
\node[d@caud, label={[dlbl]below:Config\\$\conf_{\aud}$}] at (11.2,-1) {};
\node[d@rep, label={[dlbl]below:Report\\$\AR_{\aud}$}] at (12.2,-1) {};
\node[d@hash, label={[dlbl]below:Digest\\$h_{S}$}] at (10.2,-2.75) {};
\node[d@hash, label={[dlbl]below:Digest\\$h_{\conf_{\bld}}$}] at (11.2,-2.75) {};
\node[d@hash, label={[dlbl]below:Digest\\$h_{\aux}$}] at (12.2,-2.75) {};
\node[d@rep, label={[dlbl]below:Report\\$\AR_{\mathrm{App}}$}] at (14.5,-0.5) {};

\matrix (legend) [matrix of nodes, anchor=north west, column sep=2.5mm, row sep=1mm,
    nodes={anchor=center}] at (-1.2,-3.6) {
  |[tile, fill=cudVermillion, minimum width=10mm, minimum height=5mm]| {} & proprietary &
  |[tile, fill=cudYellow, minimum width=10mm, minimum height=5mm]| {} & public &
  |[tile, fill=cudSky!30, draw=cudSky!60!black, minimum width=10mm, minimum height=5mm]| {} & TEE workload &
  |[tile, fill=cudBlue, minimum width=10mm, minimum height=5mm]| {} & RATS role \\
};

\end{tikzpicture}

\caption{Data flow of the proposed architecture.}
\Description{Data-flow diagram of the proposed architecture. A developer, acting as the reference value provider, supplies private source code and build configuration to an attestable builder, and additionally auxiliary input and a public audit configuration to an attestable auditor. The builder outputs a private build artifact to the attester, and supplies the verifier with public digests of the source code and the build configuration, the artifact measurement, and a public attestation report. The auditor supplies the verifier with public digests of the source code, the build configuration, and the auxiliary input, together with the public audit verdicts, the audit configuration, and a public attestation report. The attester sends its own attestation report to the verifier.}

\label{fig:arch}

\end{figure*}

In this section, we present our framework in full detail. \prettyref{fig:arch} shows the data flow.

\subsection{Notation}

We use the following notation: $h_X = H\left(X\right)$ denotes a hash digest of data $X$ under a fixed cryptographic hash function $H$. $m_X := \meas\left(X\right)$ denotes a measurement of artifact $X$ that an attestation report binds to $X$ (e.g. the {\MRENCLAVE} for an SGX enclave, the launch digest for a SEV-SNP or TDX VM, the file hash digest for Linux IMA). $\AR_{X}\left(d\right)$ denotes an attestation report of $X$ whose {\REPORTDATA} field binds data $d$.

The proprietary artifact is denoted $S$ (``source code,'' but the construction is content-agnostic: $S$ may be source code, LLM weight data, an application's configuration, a container policy file, \ldots). For artifacts that are not built from source---including the case where the auditor targets an executable directly (e.g. static binary analysis)---the attestable builder is omitted and the attestable auditor operates on the artifact as given; the set of applicable audits is naturally more limited than in the source-level case.

\subsection{Attestable build}

We adopt the idea of the attestable build, introduced by Hugenroth \emph{et al.} \citep{HLMB25}. The \emph{attestable builder} is responsible for linking the source code digest and the build artifact measurement. This component must either be open source and reproducibly buildable, or have its binaries made publicly available.

\emph{Input.} Source code $S$ (private); build configuration $\conf_{\bld}$ (private).

\emph{Processing.} (1) Compute hash digests $h_{S} := H\left(S\right)$ and $h_{\conf_{\bld}} := H\left(\conf_{\bld}\right)$; (2) perform the build; (3) obtain a build artifact $B$ with measurement $m_{B} := \meas\left(B\right)$; (4) issue an attestation report:
\begin{align*}
R_{\bld} & := \AR_{\bld}\left( H\left( d_{\bld} \right) \right), \\
d_{\bld} & := \left( h_{S}, h_{\conf_{\bld}}, m_{B} \right).
\end{align*}

\emph{Output.} The pre-hashed report data $d_{\bld}$ (public), the attestation report $R_{\bld}$ (public), and the build artifact $B$ (private; delivered only to the deployment infrastructure---never to the verifier).

\emph{Certification.} The report is a certificate that ``the source code with digest $h_{S}$, under some build configuration with digest $h_{\conf_{\bld}}$, yields a build artifact with measurement $m_{B}$.''

\subsection{Attestable audit (our proposal)}

The centrepiece of our architecture is the attestable audit. The \emph{attestable auditor} is responsible for linking the source code digest and the audit verdicts. As with the attestable builder, this component must also be inspectable.

\emph{Input.} (1) Source code $S$ (private); (2) build configuration $\conf_{\bld}$ (private; bound into the report so that the verifier can later link the audit to the corresponding attestable build, and available to build-dependent audits); (3) audit configuration $\conf_{\aud}$ (public; e.g. fuzzing budgets, linter rule set, AI auditor's prompts, axioms and goals for formal verification, \ldots); (4) auxiliary input $\aux$ (private; e.g. formal proofs).

\emph{Processing.} (1) Compute hash digests $h_{S}$, $h_{\aux}$, and $h_{\conf_{\bld}}$; (2) run the configured audit suite---dynamic analysis (e.g. fuzzing, sanitizers, Valgrind), static analysis (e.g. linting, taint analysis, Bandit), AI code auditing, formal methods (e.g. model checking, formal proof checking, type checking for strongly typed languages); (3) obtain (per-tool) verdicts $\verd$ (public); (4) issue an attestation report:
\begin{align*}
R_{\aud} & := \AR_{\aud}\left( H \left( d_{\aud} \right) \right), \\
d_{\aud} & := \left( h_{S}, h_{\conf_{\bld}}, h_{\aux}, \conf_{\aud}, \verd \right).
\end{align*}

\emph{Output.} The pre-hashed report data $d_{\aud}$ (public) and the attestation report $R_{\aud}$ (public).

\emph{Certification.} The report is a certificate that ``the pair of the source code with digest $h_{S}$ and the build configuration with digest $h_{\conf_{\bld}}$, when subjected to the audits configured by $\conf_{\aud}$ with auxiliary input with digest $h_{\aux}$, produces verdicts $\verd$.''

\emph{Verdict semantics.}
Our framework specifies no universal semantics to $\verd$.
It is an audit-specific structured output whose meaning is determined by the measured auditor and the public audit configuration $\conf_{\aud}$.
For example, a proof checker of formal verification may establish a theorem relative to its disclosed specification and assumptions, whereas a fuzzing verdict records that no failure was observed within the stated campaign.

\subsection{End-to-end protocol}
\label{sec:protocol}

Assume that the attestable builder and the attestable auditor are either (1) open source and reproducibly buildable, or (2) have their binaries made publicly available. This enables the verifier to reproduce their measurements $m_{\bld}$ and $m_{\aud}$ from their source code or binaries.

\emph{Step 1: Certification.} The attester's developer (the reference value provider) builds the attester's binary inside the attestable builder, and obtains the binary $B$ and its certificate $\left(d_{\bld}, R_{\bld}\right)$. The developer also audits the attester's source code $S$ using the attestable auditor, and obtains the certificate $\left(d_{\aud}, R_{\aud}\right)$.

\emph{Step 2: Deployment.} The developer then deploys $B$ inside the attester's environment.

\emph{Step 3: Certificate delivery.} The developer sends the certificates $\left(d_{\bld}, R_{\bld}\right)$ and $\left(d_{\aud}, R_{\aud}\right)$ to the verifier.

\emph{Step 4: Attesting the measurement.} The verifier and the attester perform the usual attestation procedure. The attester proves its own measurement $m$ to the verifier.

\emph{Step 5: Attesting the build.} The verifier reproduces the builder's measurement $m_{\bld}$ from its public release, verifies the builder's attestation report $R_{\bld}$ against the TEE root of trust, and checks that the {\REPORTDATA} and measurement fields are equal to $H\left(d_{\bld}\right)$ and $m_{\bld}$, respectively. This step proves that the source code with digest $h_{S}$, under some build configuration with digest $h_{\conf_{\bld}}$, yields a build artifact with measurement $m_{B}$.

\emph{Step 6: Attesting the audit.} The verifier reproduces the auditor's measurement $m_{\aud}$ from its public release, verifies the auditor's attestation report $R_{\aud}$ against the TEE root of trust, and checks that the {\REPORTDATA} and measurement fields are equal to $H\left(d_{\aud}\right)$ and $m_{\aud}$, respectively. This step proves that the source code with digest $h_{S}$, when subjected to the audits configured by $\conf_{\aud}$, produces verdicts $\verd$.

\emph{Step 7: Chaining the evidence.} The verifier checks that $d_{\bld}$ and $d_{\aud}$ share the same $h_{S}$ and $h_{\conf_{\bld}}$ and that $m_{B}$ in $d_{\bld}$ is equal to $m$. This step proves that the attestable builder and auditor are applied to the same source code $S$ with the same build configuration $\conf_{\bld}$ and that the attester is built from $\left(S, \conf_{\bld}\right)$.

\subsection{Attestable build and audit as CCPs}
\label{sec:unifying}

The attestable build and the attestable audit can be unified under a single generalised concept: the \emph{Confidential Computing Proof} (CCP) of Russinovich \emph{et al.} \citep{RusEtAl24}. A CCP is essentially TEE attestation viewed as a hardware-backed zero-knowledge proof (ZKP)---for a decidable (Turing computable) relation $\varphi\left(x, w\right)$, an open-source TEE workload $\Pi_{\varphi}$ decides $\varphi$ on public input $x$ and private witness $w$, and then emits an attestation report whose report-data field commits to the quantified statement $\exists w. \varphi\left(x, w\right)$; the report-plus-measurement is the proof object for the statement $\exists w. \varphi\left(x, w\right)$. We refer the reader to \citep{RusEtAl24} for the full formalism.

Concretely, the attestable build is a CCP for the relation ``$S$ hashes to $h_{S}$ and, under the build configuration $\conf_{\bld}$, builds to an artifact with measurement $m_{B}$,'' and the attestable audit is a CCP for the relation ``$S$, audited under $\conf_{\aud}$ with auxiliary input $\aux$, yields verdicts $\verd$.'' The full logical relations, together with a remark on how our workloads depart from a literal CCP decider, are given in \prettyref{app:ccp}.

\subsection{Security considerations}
\label{sec:security}

We briefly argue that the protocol of \prettyref{sec:protocol} achieves the goals of \prettyref{sec:model}, and make its residual assumptions explicit.

\emph{Binding (P1).} A malicious developer cannot forge $R_{\bld}$ or $R_{\aud}$ without breaking the TEE root of trust, and cannot obtain them from tampered workloads without changing the measurements $m_{\bld}$ and $m_{\aud}$, which the verifier reproduces independently from the public releases (Steps 5--6). Mix-and-match attacks---presenting an audit certificate issued for different source code, a different build configuration, or a different auxiliary input---are rejected by the cross-checks of Step 7, and presenting a binary other than the certified one yields an attested measurement $m \neq m_{B}$ in Step 4.

\emph{Non-disclosure (P2).} Beyond the public $\conf_{\aud}$ and $\verd$, every entry of $d_{\bld}$ and $d_{\aud}$ that the verifier receives is a digest, so nothing about $S$, $\conf_{\bld}$, $\aux$, or $B$ is revealed beyond what $\verd$ itself discloses---assuming, as usual for hash commitments, that the artifacts carry enough entropy to resist brute-force preimage guessing.

\emph{Selective disclosure.} The developer chooses which certificates to present: it may run many audits and disclose only favourable ones. A verdict therefore means ``this audit, under this configuration, produced these results,'' not ``no audit ever failed.'' Relying-party policies should hence prescribe the audit configurations they require, rather than merely accept whatever verdicts are presented.

\emph{Trusted computing base.} The soundness of the certificates rests on the measured workloads: the build toolchain inside the attestable builder and the audit tools inside the attestable auditor are part of the TCB, so a malicious-but-measured toolchain (in the spirit of trusting-trust attacks) defeats P1. Detecting such a toolchain is exactly what the public inspectability requirement of \prettyref{sec:model} enables.

% ---------------------------------------------------------------------
%  Examples
% ---------------------------------------------------------------------
\section{Examples}
\label{sec:examples}

The architecture of \prettyref{sec:arch} is attester/artifact-agnostic: the verifier merely cross-checks digests, so the CCPs are issued once under the issuer's TEE trust model and then applied to deployments whose own attestation technology may be entirely different. We sketch three instantiations, one for each attester form surveyed in \prettyref{sec:intro}.

\subsection{Proprietary SGX application}

The attester is a proprietary SGX enclave; the CCPs certify enclave-level properties such as statelessness, correct use of an RA-TLS handshake \citep{RATLS}, or the absence of tainted flows from confidential inputs to untrusted outputs. Building an SGX enclave image involves a heavyweight toolchain, so the attestable builder is best hosted in a CVM. Note that for SGX the measurement $m_{B}$ is the {\MRENCLAVE}; the builder therefore derives the expected {\MRENCLAVE} from the build output via the \texttt{sgx\_sign} tool of SGX-SDK. This is why $\meas\left(\cdot\right)$ is kept distinct from $H\left(\cdot\right)$. The audit platform is chosen per audit type: AI code auditing requires a CVM with a TEE-capable GPU (e.g. NVIDIA Confidential Computing), whereas lightweight static analysis may run inside SGX, as in our PoC (\prettyref{sec:poc}). At verification time, the attester presents an SGX quote proving the {\MRENCLAVE} of the running enclave, and the vendor---the party that would conventionally act as the reference value provider---ships the pre-generated CCPs instead of a bare reference {\MRENCLAVE}. The verifier runs the protocol of \prettyref{sec:arch} and concludes that the running enclave satisfies the audited properties.

\subsection{Proprietary script on an IMA-measured host}

The attester is a server-side script running on a Linux host with a TPM-backed IMA; the CCP certifies properties of the script. Because the script is deployed as-is, the attestable builder is omitted and the measurement of $S$ is simply the file hash recorded by IMA. The attester presents the IMA measurement log containing the script's entry, the PCRs to which the log is extended (typically PCR 10), and a TPM quote over those PCRs; together these prove the digest of the running script, and the vendor ships the audit CCP in place of a reference digest.

This example illustrates the decoupling most starkly: the audit CCP is issued under a TEE trust model even though the deployment platform of the attester is not necessarily a TEE. When the Linux OS itself is running within a CVM and the vTPM is located at a guest privilege level isolated from both the guest OS and the hypervisor \citep{NarEtAl23}, the deployment-side trust model aligns with the issuer's.

\subsection{Proprietary LLM behind an open-source server}

The attester is an inference server running inside a CVM; the server is fully open source, but the loaded model weights are proprietary. Here $S$ is the weight file and the CCP certifies model-level properties, e.g. quantitative scores on a prompt-injection-resistance benchmark; $\verd$ can carry such scores, since its format is arbitrary. Benchmarking requires GPU inference, so the attestable auditor runs in a CVM with a TEE-capable GPU (e.g. NVIDIA's Hopper, Blackwell and Rubin); no attestable builder is involved, as the audit certifies properties of the weights as given. At verification time, the CVM's launch measurement and the IMA measurement log cover the open-source server, whose behaviour the verifier can fully validate from source; the measured server computes the digest of the loaded weights and binds it into the attestation evidence (e.g. into the {\REPORTDATA}, an RTMR, or an IMA log). The verifier thus trusts the reported weight digest because it trusts the verified server code, and the vendor's CCP lifts that digest to the benchmarked property.

% ---------------------------------------------------------------------
%  PoC
% ---------------------------------------------------------------------
\section{Proof-of-concept implementation}
\label{sec:poc}

To demonstrate the feasibility of the attestable audit, we implemented a PoC that audits Python source code inside an SGX enclave and emits a CCP binding the audit verdict to the code digest.

\emph{TEE platform.} The enclave runs under the Gramine Library OS \citep{gramine}, which lets unmodified Python scripts run inside SGX with only a manifest describing the trusted file set; no algorithmic changes to the audit tools were required. By default, we use an SGX-capable VM\footnote{Microsoft Azure Standard\_DC1s\_v3, Intel(R) Xeon(R) Platinum 8370C, Ubuntu 24.04 LTS, Linux 6.17.0-1018-azure.} as an SGX host, with the Azure THIM service as a PCCS for retrieving/caching the attestation collateral.

\emph{Audit pipeline.} The auditor enclave takes a Python script as input $S$ and runs two analyses: a syntax check via Python's \texttt{ast} module and a security scan with Bandit \citep{bandit}.

\emph{CCP binding.} Following the construction of \prettyref{sec:arch}, the pre-hashed report data $d_{\aud}$ is the JSON object consisting of the audit verdicts $\verd$ and the SHA-384 digest of the given Python script $S$, and its SHA-384 digest is included in the {\REPORTDATA} field of the SGX quote.

\emph{Scope.} The PoC exercises the \emph{CCP--generation} side end-to-end: it produces a valid SGX quote whose {\REPORTDATA} commits to the audit result of a given Python script.

% ---------------------------------------------------------------------
%  Related Work
% ---------------------------------------------------------------------
\section{Related work}
\label{sec:related}

In this section, we position our framework against three lines of related work, each of which relates to ours in a different way. Code Transparency Services offer a complementary substrate on which our build and audit claims could be published; cryptographic ZKPs are an alternative proof mechanism that does not scale to the workloads we target; and Property-Based Attestation and Proof-Carrying Code share our goal of attesting properties rather than binary identity, while relying on different trust models and suffering from their own limitations in the proprietary setting, e.g. code disclosure and reproducible builds.

\subsection{Code transparency service}

The closest conceptual neighbour is the work of Delignat-Lavaud \emph{et al.} \citep{DLEtAl23}. They propose a Code Transparency Service (CTS) for confidential computing, which records signed claims about source releases, builds, policies, reviews, and deployed artifacts in a public append-only verifiable ledger and returns receipts for registered claims. CTS is claim-agnostic by design and can, in principle, record claims produced by code-scanning or code-verification services, including claims generated by auxiliary TEEs. In this sense, the output of an attestable audit could be registered in CTS.

Our work addresses a different layer of the problem. CTS provides a transparency and accountability substrate for storing and checking claims, but it does not define how a privacy-preserving audit claim about undisclosed source code should be generated, what such a claim means, or how it should be cryptographically linked to the artifact digest observed in remote attestation. Attestable audit fills this gap by running the audit inside a TEE and binding the resulting audit-derived properties to a source code digest. When composed with attestable build, this yields an end-to-end verification chain from the currently attested artifact digest to properties of the corresponding, undisclosed source code. Thus, CTS could be used to publish, timestamp, and make accountable the build and audit claims produced by our mechanism, whereas our contribution is the concrete TEE-backed evidence-generation and verification mechanism for such claims.

\subsection{Zero-knowledge proof for code properties}

Zero-knowledge proofs (ZKPs), including zk-SNARKs, zk-STARKs, and general-purpose zkVMs such as RISC Zero \citep{zkVM}, could, \emph{in principle}, serve our purpose by replacing CCPs directly. In practice, however, current general-purpose ZKP systems do not scale to the workloads we target. Across these schemes, the cost of proof generation and/or verification---and the size of the resulting proof---grows with the complexity of the verification logic for the statement being proved \citep{NKA25}. For statements as heavy as a software build, a fuzzing campaign, or a GPU-accelerated AI code audit, this dependence makes ZKPs impractical. Vanilla zk-SNARK and zk-STARK systems require that the verification logic be converted into arithmetic constraints, which are typically written in domain-specific languages such as Circom \citep{circom}, resulting in significant implementation costs. Furthermore, efficient zk-SNARK systems (such as Groth16 \citep{groth16}) require a multi-party trusted setup, which complicates the trust model. Although zkVMs remove the need to hand-write a verification circuit, the underlying complexity problem remains. Therefore, while ZKPs offer strong cryptographic guarantees, they are not  efficient enough for our purpose.

Our framework employs the Confidential Computing Proof (CCP) \citep{RusEtAl24}. A CCP does not cryptographically encode the entire evaluation trace. Instead, it provides hardware-rooted attestation evidence that proves the identity of applications running within the TEE and the claims generated by those applications. Soundness therefore relies on the hardware root of trust, the TEE isolation mechanism, and the correctness of the measured applications, which must be implemented so that they emit the claims only after the intended build or audit has completed successfully. The advantage is the cost profile: because existing applications run essentially unmodified inside the TEE---reusing existing build systems and audit pipelines---the asymptotic complexity of the computation is unaffected; the additional cost consists only of a constant-factor overhead imposed by the isolation mechanisms (e.g. memory encryption) plus the cost of attestation itself. This is dramatically smaller than the cost of existing ZKP systems, which grows with the complexity of the computation being proved, as discussed above.

This trade-off changes the engineering problem. Instead of rewriting the audit pipeline as a ZK-friendly circuit or zkVM program, the main requirement is to run it in a TEE and to bind its outputs to the attestation report. Process-based TEEs such as Intel SGX require SDK-specific re-implementation of existing applications but no algorithmic changes; Library OSes such as Gramine \citep{gramine} and Occlum \citep{Occlum} keep code modifications to a minimum. Confidential VMs such as AMD SEV-SNP and Intel TDX further improve compatibility by supporting largely unmodified VM workloads, and confidential GPU platforms extend this model to selected GPU-accelerated workloads. The cost of this practicality is a different trust model: unlike ZKPs, attestable audit depends on trusted hardware, vendor attestation infrastructure, TEE configuration, and the software TCB inside the measured environment. Our goal is therefore not to replace ZKPs, but to provide a practical remote-attestation-native mechanism for proving audit properties of complex, non-ZK-friendly software workloads.

\subsection{Property-based attestation and proof-carrying code}

The question of attesting \emph{properties} of code rather than mere binary identity is not new. Haldar \emph{et al.} \citep{HCF04} proposed Semantic Remote Attestation, in which a trusted virtual machine attests to high-level properties of the code it executes; Sadeghi and St\"uble \citep{SS04} spelled out the general principle of Property-Based Attestation (PBA), arguing that relying parties care about properties, not the specific binary that realises them. Subsequent work developed concrete PBA protocols with \citep{CheEtAl06} and without \citep{CheEtAl08} a Trusted Third Party (TTP). In a related line, Proof-Carrying Code (PCC) \citep{Nec97,Nec98} and its foundational variant \citep{App01} attach machine-checkable proofs of properties directly to a binary, allowing the code consumer to verify the property locally before execution. Our work shares the same end goal---conveying \emph{semantic} guarantees about a program rather than a bare hash digest.
 
\emph{Protected asset.} PBA hides \emph{which} of a set of known admissible configurations the attester runs, but the configurations themselves are inspected by the certifier \citep{CheEtAl06} or published as an agreed list \citep{CheEtAl08}; even the TTP-free protocol \citep{CheEtAl08} removes the trusted party only at the protocol layer, since establishing that each listed configuration possesses the property still presupposes that someone can inspect the code---which is not an option for proprietary code. Likewise, a PCC proof is delivered together with the binary and effectively discloses the code. Attestable audit instead protects the vendor's code confidentiality: no certificate, admissible list, or proof derived from the code's content is ever published, and the verifier sees only a digest, the audit configuration, and the verdicts.

\emph{Issuance versus application.} The attestable auditor acts as a certificate issuer for code properties: a CCP is issued once per $\left(S, \conf_{\aud}\right)$ and can then be consumed in the attestation of arbitrarily many deployments of the same code. The deployed workload need not itself be a TEE---the verifier merely cross-checks digests, so any measurement-emitting attester (a CVM, a confidential container, or an IMA-measured workload with a TPM) qualifies---while the unforgeability of the certificate rests on the TEE trust model of the issuer, independently of the deployment platform. Moreover, the issuance act itself is attested: the CCP proves that the measured audit pipeline actually ran on the code, whereas a PBA credential proves only that the TTP signed, not that any inspection took place \citep{CheEtAl06}; attestable audit is thus structurally a PBA whose TTP is replaced by a transparent, machine-checkable issuer, and whose TPM-bound presentation protocol is replaced by a plain digest cross-check. PCC sits at the opposite end of the spectrum: its proofs are self-certifying and need no issuer trust at all, but they can be consumed only by a party holding the code, and hence cannot serve as a certificate in remote attestation without disclosure,  and its property class is confined to what the proof logic can express (e.g. memory and type safety), whereas an attestable audit can certify any computable property.
 
\emph{Runtime binding.} A PBA credential certifies reference measurements, so binding a source-level property to the binary loaded at runtime requires either trusting the certifier's claim about the source-to-binary link or letting the verifier re-derive the link independently, which in practice demands a reproducible build \citep{LZ22} and disclosure of the source code. PCC needs no such link because the proof is constructed over the machine code itself, but only because the consumer holds the code, which conflicts with non-disclosure. Our attestable build provides the link as a CCP attesting $\left(S, B\right) \mapsto \left(h_{S}, m_{B}\right)$ under hardware-rooted signing, so an ordinary, non-reproducible build suffices, without disclosing $S$ or $B$.

\emph{Complementarity.} These approaches are complementary rather than competing: a PCC-style machine-checkable proof can be supplied as the private auxiliary witness $\aux$ and checked inside the attestable auditor, conveying proof-grade assurance to the verifier without disclosing either the code or the proof.

% ---------------------------------------------------------------------
%  Discussion
% ---------------------------------------------------------------------
\section{Discussion}
\label{sec:discussion}

\subsection{Hosting the attestable services}

Our architecture is agnostic to who operates the TEE platform on which the attestable builder and auditor run.
Under the standard TEE threat model of \prettyref{sec:model}, all software and operators outside the TEE are untrusted, so the guarantees formally hold regardless of the host.
In practice, however, this model is strained by low-cost physical attacks. TEE.fail \citep{TeeFail} demonstrates that attestation signing keys of current TEE platforms can be extracted via DDR5 memory-bus interposition with equipment costing under a thousand dollars, enabling the forgery of attestation reports.

Given such attacks, a situation in which a party that has an incentive to cheat also has physical access to the hosting machine is undesirable, even though it is nominally covered by the threat model.
Therefore, the TEE platform serving the attestable builder and auditor should be provided by a neutral third party without incentive to attack either side---neither to forge certificates as a self-hosting developer could, nor to harvest the developer's proprietary inputs as a malicious host could.

Cloud service providers are natural candidates: if a CSP qualifies as such a party, using its TEE platform is a particularly desirable choice, because CSP platforms are isolated inside tightly controlled data centres---whose physical locations are often even undisclosed---making physical access by any interested party difficult.

\subsection{Certificate lifecycle and TCB recovery}

The certificates (attestation reports) issued by the attestable builder and auditor bind the digest of $S$ to the result of a completed computation.
The build report certifies the resulting artifact measurement, whereas the audit report certifies the resulting verdicts.
As specified in \prettyref{sec:protocol}, these certificates are generated before deployment and reused when verifying deployments of the certified artifact.
Their role is therefore not to establish the freshness of a particular verification session.
Consequently, they require no session binding and need not include a nonce.

This reuse assumes that the TCB under which a certificate was issued remains trusted.
If a vulnerability or compromise triggers TCB recovery, the platform updates its components (such as microcode and firmware), and the endorser revokes the affected TCB version through updated endorsements.
A certificate issued under the revoked TCB then fails verification and must be reissued by rerunning the corresponding computation on the recovered platform.

For the attestable builder, reissuance requires rebuilding $S$.
If the rebuilt artifact has a measurement different from that of the running attester, the check in Step~7 fails and the artifact must be redeployed.
A reproducible build avoids this cost by reproducing the same artifact measurement, allowing only the build certificate to be reissued.

\subsection{Dependency coverage in auditing}

The dependencies covered by an audit depend on its analysis.
An audit confined to source $S$ need not resolve dependencies, whereas a dependency-aware audit may inspect direct and transitive dependencies.
In the latter case, their resolution must match that used to build artifacts $B$.
Pinning the dependencies and using a reproducible build can make this correspondence deterministic.

Reproducibility of $B$, however, does not fix dependencies supplied at run time.
A dynamically linked library installed in the execution environment may change without changing $B$.
This is a limitation of auditing only the application artifact.
To address this issue, code outside that artifact must be bound separately by platform-side attestation.
For example, CCPs for a confidential container may cover its container image, while CVM attestation must separately establish the expected container runtime and policy engine.

% ---------------------------------------------------------------------
%  Conclusion
% ---------------------------------------------------------------------
\section{Conclusion}
\label{sec:concl}

We identified a fundamental gap in remote attestation for proprietary software: matching a measurement against a reference value tells a verifier only that ``something with this digest is running,'' forcing full trust in the reference value provider. We proposed an architecture that closes this gap by composing an attestable build and an attestable audit, both cast as Confidential Computing Proofs: open-source TEE workloads that bind a source code digest to a build artifact digest and to an audit verdict without disclosing either the source code or the build artifact to the verifier. This enables property-based attestation for proprietary applications without code disclosure. A PoC on Intel SGX with the Gramine Library OS, auditing Python code with Bandit, demonstrates that CCP generation is practical with off-the-shelf tools.

\subsection*{Acknowledgements}

The author is grateful to the anonymous reviewers for their helpful comments.

\bibliography{refs.bib}

\appendix

\section{Open Science}

\url{https://github.com/acompany-develop/attestable-audit-poc}

\section{CCP Relations for the Attestable Build and Audit}
\label{app:ccp}

This appendix gives the concrete CCP relations $\varphi\left(x, w\right)$ underlying the CCP view of \prettyref{sec:unifying}, where $x$ is the public input and $w$ is the private witness.
The attestable build can be considered as a CCP $\Pi_{\varphi_{\bld}}$ for the following relation:
\begin{align*}
  x &= \left(h_{S}, h_{\conf_{\bld}}, m_{B}\right), \\
  w &= \left(S, B, \conf_{\bld}\right), \\
  \varphi_{\bld}(x, w) &\equiv \big[ h_{S} = H\left(S\right) \land m_{B} = \meas\left( B \right)\\
        &\qquad \land \mathtt{Build}(S, \conf_{\bld}) = B \big].
\end{align*}
The attestable audit can be considered as a CCP $\Pi_{\varphi_{\aud}}$ for the following relation:
\begin{align*}
  x &= \left(h_{S}, \conf_{\aud}, h_{\conf_{\bld}}, h_{\aux}, \verd \right), \\
  w &= \left(S, \conf_{\bld}, \aux \right), \\
  \varphi_{\aud}(x, w) &\equiv \big[ h_{S} = H\left(S\right) \land h_{\conf_{\bld}} = H \left( \conf_{\bld} \right) \\
        & \qquad \land h_{\aux} = H\left(\aux\right) \\
        & \qquad \land \mathtt{Audit}(S, \conf_{\bld}, \conf_{\aud}, \aux) \\
        & \qquad = \verd \big].
\end{align*}

Note that the attestable build and audit optimise the literal CCP model in two ways.
First, rather than taking the build artifact $B$ (and hence its measurement $m_{B}$) or the audit verdicts $\verd$ as externally computed inputs, they compute these values inside the TEEs.
This avoids the duplication of computing these values when preparing inputs and issuing CCPs.
Second, the attestable builder not only issues a CCP but also also outputs $B$.
The artifact produced by the CCP issuer (attestable builder) can then be deployed directly rather than rebuilt separately.
These optimisations eliminate redundant computation while yielding the same proof object.

\end{document}